\documentclass[hyper]{JHEP} % 10pt is ignored!

\usepackage{epsfig}
\newcommand\fverb{\setbox\pippobox=\hbox\bgroup\verb}

\newcommand\fverbdo{\egroup\medskip\noindent%

            \fbox{\unhbox\pippobox}\ }

\newcommand\fverbit{\egroup\item[\fbox{\unhbox\pippobox}]}

\newbox\pippobox

\title{Stable and Unstable Dp-branes in $p-$ brane
Newton-Cartan Background}
\author{J. Kluso\v{n}\\
Department of
Theoretical Physics and Astrophysics\\
Faculty of Science, Masaryk University\\
Kotl\'{a}\v{r}sk\'{a} 2, 611 37, Brno\\
Czech Republic\\
E-mail: \email{klu@physics.muni.cz}} \preprint{}

 \abstract{We formulate p-brane Newton-Cartan background through the limiting
 procedure from relativistic Dirac-Born-Infeld action and Wess-Zumino term. We also determine action for unstable D(p+1)-brane
in p-brane Newton-Cartan Background and study its properties.}
\def\hB{\hat{B}}
\def\tba{\tilde{\ba}}
\def\tl{\tilde{l}}

\def\tE{\tilde{E}}

\def\hphi{\hat{\phi}}

\def\hgamma{\hat{\gamma}}
\def\homega{\hat{\omega}}

\def\bA{\mathbf{A}}
\def\hE{\hat{E}}

\def\ttau{\tilde{\tau}}
\def\hbt{\hat{\bt}}

\def\hE{\hat{E}}
\def\tF{\tilde{F}}

\def\tT{\tilde{T}}

\def\hG{\hat{G}}

\def\bt{\mathbf{t}}
\def\tbt{\tilde{\bt}}

\def\be{\begin{equation}}

\def\halpha{\hat{\alpha}}
\def\hbeta{\hat{\beta}}
\def\ee{\end{equation}}

\def\bea{\begin{eqnarray}}
\def\bh{\bar{h}}

\def\eea{\end{eqnarray}}

\def\hdelta{\hat{\delta}}
\def\homega{\hat{\omega}}

\def\bX{\mathbf{X}}

\def\mH{\mathcal{H}}

\def\tA{\tilde{A}}

\newcommand{\hC}{\hat{C}}

\def\tC{\tilde{C}}

\newcommand{\tB}{\tilde{B}}

\newcommand{\hA}{\hat{A}}
\newcommand{\mF}{\mathcal{F}}

\newcommand{\hD}{\hat{D}}

\def \bA{\mathbf{A}}

\newcommand{\ba}{\mathbf{a}}
\newcommand{\tphi}{\tilde{\phi}}

\begin{document}
%%%%%%%%%%%%%%%%%%%%%
%%%%Introduction %%%%%%%%%
%%%%%%%%%%%%%%%%%%%%
\section{Introduction and Summary }
In the past few years there was renewed interest in the study of non-relativistic
theories especially in Newton-Cartan (NC) formulation. NC geometry was introduced
by Cartan \cite{Cartan:1923zea} as to geometrise Newton's laws of gravitation but last few years showed that this formalism can be used in different areas of modern physics and it can be also extended in many ways. For example,  it was shown that
torsional generalization of NC geometry which possesses non-closed clock form
was observed as boundary geometry in the context of Lifschitz holography
\cite{Christensen:2013lma,Christensen:2013rfa,Hartong:2014oma}. This geometry is known
as type I torsional NC geometry in terminology \cite{Hansen:2020pqs,Hansen:2019svu}. This geometry also plays an important role in the context of the non-relativistic string theory
in torsional NC background \cite{Kluson:2020aoq,Roychowdhury:2020kma,Kluson:2019ajy,Pereniguez:2019eoq,Kluson:2019avy,Harmark:2019upf,Harmark:2018cdl,Kluson:2018egd}. However there is another version of non-relativistic string theory in stringy NC background
\cite{Andringa:2012uz,Kluson:2018uss,Bergshoeff:2019pij,Kluson:2019uza,Gomis:2019zyu,
Kluson:2018vfd,Roychowdhury:2019qmp,Kluson:2019ifd,Kluson:2018grx}. This theory is defined by splitting target space dimensions into longitudinal and transverse ones respectively. Say differently, while natural probe object in NC background is point particle in case of the stringy NC theory this object is fundamental string as was firstly shown in
\cite{Andringa:2012uz}. It was also argued there that this procedure can be naturally generalized to the p+1 -dimensional object known as p-brane.   Analysis of this proposal  was performed in
\cite{Kluson:2017abm,Roychowdhury:2019qmp,Kluson:2019uza}. On the other hand it is well known  that extended objects in string theories are Dp-branes
 \footnote{For review and extensive list of references, see
\cite{Simon:2011rw}.} rather than ordinary p-branes.  These $p+1$ dimensional objects are more general than $p-$brane since they couple to Ramond Ramond (RR) forms and to Neveu-Schwarz form and also there is a gauge field that propagates on their world-volume so that non-relativistic limit that defines
corresponding $p-$brane NC geometry is more difficult to implement. The goal of this paper is exactly to do this procedure and also to extend it to the case of unstable D(p+1)-branes.

Let us be more explicit. We consider Dp-brane in general background with non-trivial metric, NSNS two form and RR fields together with dilaton. Then we generalize limiting procedure introduced  in \cite{Bergshoeff:2019pij} to the case of $p+1$-dimensional probe when we split target space coordinates to $p+1-$longitudinal ones and $9-p$ transverse ones and choose corresponding form of the background metric and RR $p+1$-form.
We should also choose appropriate scaling of the string length and NSNS two forms together
with lower dimensional RR forms so that non-relativistic D-brane has finite coupling
to these fields. As a result we obtain an action for non-relativistic Dp-brane in p-brane NC background. As a check we show that in the special case $p=2$
this non-relativistic D2-brane agrees with the action that arises by dimensional reduction of non-relativistic M2-brane that was analysed in \cite{Kluson:2019uza}.

As a next step in our analysis we focus on non-relativistic unstable D(p+1)-brane. In the process of its definition we split target space coordinates into $p+1-$longitudinal ones as in case of the stable $p+1-$brane since we treat $T$ as an additional embedding coordinate following previous work \cite{Erkal:2009xq}. This is crucial difference with respect to the paper \cite{Roychowdhury:2019qmp} where the number of longitudinal dimensions coincide with the number of world-volume dimensions of unstable Dp-brane. We also consider more general case with non-trivial world-volume gauge field whose field strength has to be multiplied wish string length from dimensional reason. As a result we get unstable D(p+1)-brane in general background. As a check of our approach we study tachyon kink solution on its world-volume \cite{Sen:2003tm}
and we find that it corresponds to non-relativistic Dp-brane which is nice consistency check. We further analyse this unstable D(p+1)-brane at the tachyon vacuum corresponding to the case when the tachyon potential vanishes. This problem however is not well defined at the Lagrangian level where Lagrangian vanishes and hence it is more natural to switch to the
Hamiltonian formalism \cite{Sen:2000kd,Sen:2003bc}. It was shown in these papers that in case of relativistic unstable Dp-brane tachyon vacuum solution corresponds to the gas of fundamental strings. We choose similar strategy in case of non-relativistic unstable D(p+1)-brane when
we firstly determine corresponding Hamiltonian. Then we study the tachyon vacuum solution and we find not very satisfactory result that the Hamiltonian density does not correspond to non-relativistic fundamental string. This fact has natural explanation when we take into account that non-relativistic fundamental string is defined with the specific form of NSNS two form that defines stringy NC geometry while in case of p-brane NC geometry this NSNS two
form is not specified.

Let us outline our results and suggest possible extension of our work. We found non-relativistic Dp-brane action in p-brane NC gravity. We further derived an action for unstable D(p+1)-brane and we analysed tachyon condensation on its word volume and we showed
that it results to stable Dp-brane. We also derived Hamiltonian for this unstable D(p+1)-brane
and we analysed its behaviour at the tachyon vacuum state. We found that the resulting Hamiltonian density does not correspond to non-relativistic fundamental string which differs crucially from the tachyon condensation in case of relativistic unstable D(p+1)-brane.
It would be certainly nice to understand this problem better and we mean that possible resolution can be found when we take more general form of the limiting procedure with non-trivial ansatz for  NSNS two form as well. This problem is currently under investigation.

This paper is organized as follows. In the next section (\ref{second}) we find an action for non-relativistic Dp-brane in p-brane NC background. Then in section (\ref{third}) we find similar action for unstable D(p+1)-brane and study the tachyon kink solution on its worldvolume. In section (\ref{fourth}) we derive Hamiltonian form of this action and  study
tachyon vacuum solution.

\section{Non-Relativistic Dp-brane action in p-brane Newton-Cartan Background}
\label{second}
In this section we generalize our analysis of definition of M-brane non-relativistic
action to the case of Dp-brane as a probe. Explicitly, we start with the action for
Dp-brae in string theory that has the form
\begin{equation}
S=-\ttau_{p}\int d^{p+1}\xi e^{-\hphi}\sqrt{-\det \bA_{\alpha\beta}}
+\ttau_{p}\int \sum_i \hC^{(i)} \wedge e^{l_s^2 \tF+\hB} ,
\end{equation}
where $\ttau_{p}=\frac{1}{\tl_s^{p+1}}$ is $Dp-$brane tension and where $\tphi$ is dilaton. The WZ term can be expanded as
\begin{eqnarray}
& &\int \sum_i \hC^{(i)} \wedge e^{l_s^2 \tF+\hB}=\nonumber \\
& &=\int \hC^{(p+1)}+\int \hC^{(p-1)}\wedge
(l_s^2 \tF+\hB)+\frac{1}{2}\int \hC^{(p-3)}\wedge (l_s^2\tF+\hB)\wedge (l_s^2 \tF+\hB)+\dots \ ,
\nonumber \\
\end{eqnarray}
 and where $\hC^{(p+1)}$ is a pull-back of $p+1$-form to the world-volume of Dp-brane
\begin{eqnarray}
\hC^{(p+1)}&=&\hC_{\mu_1 \mu_2\dots \mu_{p+1}}dx^{\mu_1}\wedge dx^{\mu_2}\wedge \dots \wedge  dx^{\mu_{p+1}}=\nonumber \\
&=&\frac{1}{(p+1)!}\epsilon^{\alpha_1\alpha_2\dots\alpha_{p+1}}
\hC_{\mu_1 \mu_2 \dots \mu_{p+1}}\partial_{\alpha_1}x^{\mu_1}\partial_{\alpha_2}x^{\mu_2} \dots\partial_{\alpha_{p+1}}x^{\mu_{p+1}} \ .
\end{eqnarray}
Further, $x^\mu(\xi),\mu,\nu=0,1,\dots,9$ are world-volume fields that parameterise an embedding of Dp-brane in target space-time with the metric $G_{\mu\nu}$ so that $(p+1)\times (p+1)$ matrix $\bA_{\alpha\beta}$ has the form
\begin{equation}
\bA_{\alpha\beta}=G_{\mu\nu}\partial_\alpha x^\mu \partial_\beta x^\nu+B_{\mu\nu}
\partial_\alpha x^\mu\partial_\beta x^\nu+\tl_s^2 \tF_{\mu\nu} \ ,
\end{equation}
where $\partial_\alpha\equiv \frac{\partial}{\partial \xi^\alpha}$ and where $\xi^\alpha \ , \alpha,\beta=0,1,\dots,p$ parameterise world-volume of Dp-brane.

With the help of this action we proceed to the definition of non-relativistic limit, following
 \cite{Bergshoeff:2015uaa} and \cite{Bergshoeff:2019pij}. The starting
point is to express metric $G_{\mu\nu}$ with the help of the vierbein
 $\hE_\mu^{ \ \hA}$ so that
\begin{equation}\label{defG}
G_{\mu\nu}=\hE_M^{ \ \hA}\hE_\nu^{ \ \hB}\eta_{\hA\hB} \ , \quad  \hE_\mu^{ \ \hA}\hE^{\mu}_{ \ \hB}=
\delta^{\hA}_{ \ \hB} \ , \quad  \hE^\mu_{ \ \hA}\hE_\nu^{ \ \hA}=\delta^\mu_\nu \ .
\end{equation}
We define p-brane Newton-Cartan background as the generalization of stringy Newton-Cartan  background. Explicitly,
 we   split target-space indices $\hA$ into $\hA=(A,A')$ where now $A=0,1,,\dots,p$ and $A'=p+1,\dots,9$.
Then we introduce following parametrization of relativistic vierbein
\cite{Bergshoeff:2019pij}
\begin{equation}\label{relvierin}
\hE_\mu^{ \ A}=X_\mu^{ \ A}+\frac{1}{\omega}m_\mu^{ \ A} \ , \quad
 X_\mu^{ \ A}=\omega \tau_\mu^{ \ A}-\frac{1}{\omega}C_\mu^{ \  A} \ , \quad
\hE_\mu^{ \ A'} =E_\mu^{ \ A'} \ ,
\end{equation}
where $\omega$ is free parameter that we take to infinity when we define non-relativistic limit. Further, $C_\mu^{ \ A}$ are arbitrary functions which do not show up in the geometrical objects of the brane Newton-Cartan geometry that will originate
from the limiting procedure as was shown in case of stringy NC geometry in
\cite{Bergshoeff:2019pij}. Further, using (\ref{defG}) and taking the limit $
\omega\rightarrow \infty$ we obtain following important relations
\begin{eqnarray}
E_\mu^{ \ A'}E^\mu_{ \ B'}=\delta^{A'}_{B'} \ ,  \quad
E_\mu^{ \ A'}E^\nu_{ \ A'}=\delta_\mu^\nu-\tau_\mu^{ \ A}
\tau^\nu_{ \ A} \ , \quad \tau^\mu_{ \ A}\tau_\mu^{ \ B}=\delta_A^B \ , \quad
\tau^\mu_{ \ A}E_\mu^{ \ A'}=0 \ , \quad
\tau_\mu^{ \ A}E^\mu_{ \ A'}=0 \ . \nonumber \\
\end{eqnarray}
%Note that the inverse vierbein to (\ref{relvier}) has the form  (up to terms of order $\omega^{-3}$)
%\begin{equation}\label{relvierinv}
%\hE^\mu_{ \ A}=\frac{1}{\omega}\tau^\mu_{ \ A}-\frac{1}{\omega^3}\tau^\mu_{ \ B}
%(m_\nu^{ \ B}-C_\nu^{ \ B})
%\tau^\nu_{ \ A} \ , \quad \hE^\mu_{ \ A'}=E^\mu_{ \ A'}
%-\frac{1}{\omega^2} \tau^\mu_{ \ A}(m_\nu^{ \ A}-C_\nu^{ \ A})E^\nu_{ \ A'} \ .
%\end{equation}
Then with the help of (\ref{relvierin})  we obtain following form of the metric
\begin{eqnarray}\label{hGmetric}
\hG_{\mu\nu}&=&\hE_\mu^{ \ A}\hE_\nu^{ \ B}\eta_{AB}+\hE_\mu^{ \ A'}\hE_\nu^{ \ B'}\delta_{A'B'}
=\nonumber \\
&=&\omega^2 \tau_{
\mu\nu}+\bh_{MN}-(\tau_\mu^{ \ A}C_\nu^{ \ B}+\tau_\nu^{ \ A}
C_\mu^{ \ B})\eta_{AB}+\frac{1}{\omega^2}(m_\mu^{ \ A}-C_\mu^{ \ A})
(m_\nu^{ \ B}-C_\nu^{ \ B})\eta_{AB} \ , \nonumber \\
%G^{MN}&=&E^M_{ \ a}E^N_{ \ b}\eta^{ab}+E^M_{ \ a'}E^N_{ \ b'}\delta^{a' b'}=
%\nonumber \\
%&=& \frac{1}{\omega^2}\tau^{MN}+h^{MN}
%-\frac{1}{2\omega^2}(\tau^N_{ \ b}m_K^{ \ b}h^{KM}
%+\tau^M_{ \ b}m_K^{ \ b}h^{KN})
%-\nonumber \\
%&-&\frac{1}{2\omega^4}
%(\tau^M_{ \ c}m_{K}^{ \ c}\tau^{KN}+
%\tau^N_{ \ d}m^{ \ d}_{K}\tau^{KM})
%+\frac{1}{4\omega^4}\tau^M_{ \ a}m_K^{ \ a}
%h^{KL}\tau^N_{ \ b}m_L^{\ b} \ . \nonumber \\
\end{eqnarray}
where we defined
\begin{equation}
\bh_{\mu\nu}=h_{\mu\nu}+\tau_\mu^{ \ A}m_\nu^{ \ B}\eta_{AB}+
m_\mu^{ \ A}\tau_\nu^{ \ B}\eta_{AB} \  , \quad
\tau_{\mu\nu}=\tau_\mu^{ \ A}\tau_\nu^{ \ B}
\eta_{AB} \ .
\end{equation}
As the next step we  presume that Ramond-Ramond fields  have following
form and scaling
\begin{eqnarray}\label{hCRR}
\hC^{(p+1)}_{\mu_1\dots\mu_{p+1}}=e^{-\hphi}X_{\mu_1}^{ \ A_1}\dots X_{\mu_{p+1}}^{\ A_{p+1}}
\epsilon_{A_1\dots A_{p+1}}+\omega^{k_{p+1}}C^{(p+1)}_{\mu_1\dots \mu_{p+1}} \ , \quad
\hC^{(p-1)}_{\mu_1\dots \mu_{p-1}}=\omega^{k_{p-1}}C^{(p-1)}_{\mu_1\dots \mu_{p-1}} \ ,
\nonumber \\
\end{eqnarray}
where $k_{p+1},k_{p-1},\dots$ will be determined from the requirement that these
terms give finite contribution to the resulting non-relativistic action.

Finally we will presume that   NSNS two form, world-volume gauge field and string length scale as
\begin{equation}\label{hB}
\hB_{\mu\nu}=\omega B_{\mu\nu} \ , \quad  \tA_\mu=\frac{1}{\omega} A_\mu \ ,
\quad \tl_s=\omega l_s \ .
\nonumber \\
\end{equation}

Now the object $\tau_\alpha^{ \ A}=\tau_\mu^{ \ A}\partial_\alpha x^\mu$ is $(p+1)\times (p+1) $ matrix that in adapted coordinates is equal to $\tau_\alpha^{ \ A}=\mathrm{diag}(1,\dots,1)$. Hence it is natural to presume that the matrix $\tau_\alpha^{ \ A}$ is non-singular so that $\ba_{\alpha\beta}=\tau_\alpha^{ \ A}\tau_\beta^{ \ B}\eta_{AB}$ is non-singular too. Then we can introduce inverse $\tba^{\alpha\beta}$ that obeys
\begin{equation}
\ba_{\alpha\beta}\tba^{\beta\gamma}=\delta_\alpha^{ \gamma} \ .
\end{equation}
Then we easily obtain
\begin{eqnarray}
& &\ttau_p e^{-\hphi}\sqrt{-\det \bA_{\alpha\beta}}=\nonumber \\
%=\frac{1}{\omega^{p+1}}\tau_p e^{-\hphi}
%\left(\det (\omega^2 \ba_{\alpha\beta}+
%	\bh_{\alpha\beta}-(\tau_\alpha^{ \ A}C_\beta^{ \ B}+\tau_\beta^{ \ A}
%	C_\alpha^{ \ B})\eta_{AB}+\frac{1}{\omega^2}(m_\alpha^{ \ A}-C_\alpha^{ \ A})
%	(m_\beta^{ \ B}-C_\beta^{ \ B})\eta_{AB}+\right.\nonumber \\
%	+\left. \omega (l_s^2F_{\alpha\beta}+B_{\alpha\beta})\right)^{1/2}=\tau_pe^{-\hphi}\sqrt{-\det \ba_{\alpha\beta}}\times \nonumber \\
%\times
%\left(\det (\delta^\alpha_\beta+
%	\frac{1}{\omega^2}\tba^{\alpha\gamma}\bh_{\gamma\beta}
%-\frac{1}{\omega^2}\ba^{\alpha\gamma}(\tau_\gamma^{ \ A}C_\beta^{ \ B}+
%\tau_\gamma^{ \ A}C_\beta^{ \ B})\eta_{AB}+	
%	\frac{1}{\omega^4}\ba^{\alpha\gamma}(m_\gamma^{ \ A}-C_\gamma^{ \ A})
%	(m_\beta^{ \ B}-C_\beta^{ \ B})\eta_{AB}\right.+\nonumber \\
%\left.+\frac{1}{\omega} \ba^{\alpha\gamma}(l_s^2F_{\gamma\beta}+B_{\alpha\beta})\right)^{1/2}	
& &=\tau_p e^{-\hphi}\sqrt{-\det\ba_{\alpha\beta}}\left(1+\frac{1}{2\omega^2}\tba^{\alpha\beta}\bh_{\alpha\beta}-
 \frac{1}{2\omega^2}\tba^{\alpha\gamma}(\tau_\gamma^{ \ A}C_\alpha^{ \ B}+
 \tau_\gamma^{ \ A}C_\alpha^{ \ B})\eta_{AB} -\right.
\nonumber \\
& &\left.-\frac{1}{4\omega^2}\tba^{\alpha\gamma}(l_s^2F_{\gamma\delta}+B_{\gamma\delta})
\tba^{\delta\omega}(l_s^2 F_{\omega\alpha}+B_{\omega\alpha})\right) \ .
\nonumber \\
\end{eqnarray}
As we argued above the matrix $\tau_\alpha^{ \ A}$ is non-singular with inverse matrix $\tau^{\alpha}_{ \ A}$ that obeys
\begin{equation}
\tau_\alpha^{ \ A}\tau^\beta_{ \ A}=\delta_\alpha^{ \ \beta} \ , \quad
\tau_\alpha^{ \ A}\tau^\alpha_{ \ B}=\delta^A_B \ .
\end{equation}
Hence $\tba^{\alpha\beta}=\tau^\alpha_{ \ A}\tau^\beta_{ \ B}\eta^{AB}$ and
hence
\begin{equation}
\tba^{\alpha\beta}\tau_\beta^{ \ A}=\tau^\alpha_{ \ C}\eta^{CA} \ .
\end{equation}
%so that we can finally write
%\begin{eqnarray}
%\ttau_p e^{-\hphi}\sqrt{-\det \bA_{\alpha\beta}}=
%\tau_p e^{-\hphi}\det \tau_\alpha^{ \ a}
%+\frac{\tau_p}{2\omega^2}e^{-\hphi}\tau_p \sqrt{-\det\ba}\ba^{\alpha\beta}\bh_{\beta\alpha}
% -\frac{\tau_p}{\omega^2}e^{-\hphi}\det\tau_\alpha^{ \ A}\tau^\alpha_{ \ A}C_\alpha^{ \ A}
%\nonumber \\
%\end{eqnarray}
As the next step we analyse contribution from the coupling to RR field
\begin{eqnarray}
& &\ttau_p\hC^{(p+1)}=\frac{1}{\omega^{p+1}}\tau_p e^{-\hphi}\frac{1}{(p+1)!}\epsilon^{\alpha_1\dots \alpha_{p+1}}X_{\alpha_1}^{ \ A_1}\dots X_{\alpha_{p+1}}^{\ A_{p+1}}
\epsilon_{A_1\dots A_{p+1}}+\frac{1}{\omega^{p+1}}\tau_p\omega^{k_{p+1}}C^{(p+1)}= \nonumber \\
& &=\frac{1}{\omega^{p+1}}\tau_p e^{-\hphi}\det X_\alpha^{ \ A}+\omega^{k_{p+1}-(p+1)}\tau_p C^{(p+1)}=\nonumber \\
& &=\tau_p e^{-\hphi}\det  \tau_\alpha^{ \ B}\det(\delta_B^{ \ A}-\frac{1}{\omega^2}\tau_B^{ \ \gamma}C_\gamma^{ \ A})+\omega^{k_{p+1}-(p+1)}\tau_p C^{(p+1)}=\nonumber \\
& &=\tau_p e^{-\hphi}\det \tau_\alpha^{ \ A}(1-\frac{1}{\omega^2}\tau_A^{ \ \alpha}C_\alpha^{ \ A})++\omega^{k_{p+1}-(p+1)}\tau_p C^{(p+1)} \ , \nonumber \\
\end{eqnarray}
where $X_\alpha^{ \ A}=X_\mu^{ \ A}\partial_\alpha x^\mu$.
We see that in order to have finite WZ term in the action we have to impose condition
\begin{equation}
k_{p+1}=p+1 \ .
\end{equation}
In case of the lower dimensional RR fields we find, taking into account expansion of
the exponential function and the fact that $\tl_s^2 \tF_{\alpha\beta}+\hB_{\alpha\beta}=
\omega (l_s^2 F_{\alpha\beta}+B_{\alpha\beta})$ we obtain
\begin{equation}
k_{p-1}=p \ , k_{p-3}=p-1  \ , \dots
\end{equation}
finally we see that in order to keep finite part of the action we have to have following scaling of dilaton in the form
\begin{equation}
%e^{-\hphi}=\omega^2 e^{-\phi} \Rightarrow
 \hphi=\phi-2\ln \omega
\ .
\end{equation}
Collecting these results we obtain an action for non-relativistic Dp-brane in the form
\begin{eqnarray}\label{Sactnon}
& &S=-\tau_p\int d^{p+1}\xi
\sqrt{-\det\ba}e^{-\phi}\left[\frac{1}{2}\tba^{\alpha\beta}\bh_{\alpha\beta}
-\frac{1}{4}\tba^{\alpha\gamma}(l_s^2F_{\gamma\delta}+B_{\gamma\delta})
\tba^{\delta\omega}(l_s^2 F_{\omega\alpha}+B_{\omega\alpha})\right]
+\nonumber \\
& &+\tau_p \int d^{p+1}\xi (C^{(p+1)}+C^{(p-1)}\wedge (l_s^2 F+B)+\dots) \ .
\nonumber \\
\end{eqnarray}
Note that for the special case $p=2$ this action coincides with non-relativistic
D2-brane action that was derived with the help of the dimensional reduction
of non-relativistic M2-brane in
\cite{Kluson:2019uza}.

\section{Non-Relativistic Unstable D(p+1)-brane}\label{third}
In this section we find an action for non-relativistic unstable D(p+1)-brane in
stringy NC background by appropriate limiting procedure. Recall that the tachyon effective
action has the form
\footnote{For review, see\cite{Sen:2004nf}.}
\cite{Sen:1999md,Bergshoeff:2000dq,Kluson:2000iy}
\begin{equation}
S_{NBPS}=-\ttau_{p+1}^{unst}\int d^{p+2}\xi
e^{-\hphi}V(T)\sqrt{-\det\bA}+\ttau^{uns}_{p+1} \int \sum_i V(T) \tl_s dT\wedge\tC^{(i)}\wedge e^{l_s^2 \tF+\tB} \ ,
\end{equation}
where $T$ is tachyon that propagates on the world-volume of unstable D(p+1)-brane whose kinetic term is contained in the matrix $\bA$
\begin{equation}
\bA_{\alpha\beta}=G_{\mu\nu}\partial_\alpha x^\mu\partial_\beta x^\nu+
\tl_s^2\partial_\alpha T\partial_\beta T+\tl_s^2 F_{\alpha\beta}+\hB_{\mu\nu}
\partial_\alpha x^\mu\partial_\beta x^\nu  \ .
\end{equation}
%\begin{equation}
%G_{\mu\nu}=\hE_M^{ \ \hA}\hE_\nu^{ \ \hB}\eta_{\hA\hB} \ , \quad  \hE_\mu^{ \ \hA}\hE^{\mu}_{ \ \hB}=
%\delta^{\hA}_{ \ \hB} \ , \quad  \hE^\mu_{ \ \hA}\hE_\nu^{ \ \hA}=\delta^\mu_\nu \ .
%\end{equation}
Finally $V(T)$ is tachyon potential that is even function of $T$ with the minimum $V(T_{min})=0$ for $T_{min}=\pm \infty$ and $V(T_{max})=1$ for $T_{max}=0$.

For reasons that will be clear from discussion below we split target-space indices
in the same way as in case of stable Dp-brane, explicitly we   split target-space indices $\hA$ into $\hA=(A,A')$ where now $A=0,1,\dots,p$ and $A'=p+1,\dots,9$.
%Then we introduce $\tau_\mu^{ \ A}$ so that we write
%\begin{equation}
%\tau_{\mu\nu}=\tau_\mu^{ \ A}\tau_\nu^{ \ B}
%\eta_{AB} \ , a,b=0,1,,\dots,p \ .
%\end{equation}
%In the same way we introduce vierbein $E_\mu^{ \ A'}, A'=p+1,\dots,9$ and also  introduce gauge field  $m_\mu^{ \ A}$. The $\tau_\mu^{ \ A}$ can be interpreted as the gauge fields of the longitudinal translations while $E_M^{ \ A'}$  as the gauge fields of the transverse translations
%\cite{Andringa:2012uz}. Then we can also introduce their inverses with respect to their longitudinal and transverse translations
%\begin{eqnarray}
%E_\mu^{ \ A'}E^\mu_{ \ B'}=\delta^{A'}_{B'} \ ,  \quad
%E_\mu^{ \ A'}E^\nu_{ \ A'}=\delta_\mu^\nu-\tau_\mu^{ \ A}
%\tau^\nu_{ \ A} \ , \quad \tau^\mu_{ \ A}\tau_\mu^{ \ B}=\delta_A^B \ , \quad
%\tau^\mu_{ \ A}e_\mu^{ \ A'}=0 \ , \quad
%\tau_\mu^{ \ A}e^\mu_{ \ A'}=0 \ . \nonumber \\
%\end{eqnarray}
Then we introduce following parametrization of relativistic vierbein as in (\ref{relvierin})
with the metric given in
%\cite{Bergshoeff:2015uaa}
%\begin{equation}\label{relvier}
%\hE_\mu^{ \ A}=X_\mu^{ \ A}+\frac{1}{\omega}m_\mu^{ \ A} \ , X_\mu^{ \ A}=\omega \tau_\mu^{ \ A}-\frac{1}{\omega}C_\mu^{ \  A} \ , \quad
%\hE_\mu^{ \ A'} =E_\mu^{ \ A'} \ ,
%\end{equation}
%where $\omega$ is free parameter that we take to infinity when we define non-relativistic limit.
(\ref{hGmetric}).  Further, RR fields, NSNS two forms and scaling of remaining world-volume
fields are given in (\ref{hCRR}) and (\ref{hB}). Note due to the fact that unstable D(p+1)-brane contains potential which is function of $T$ and we want this function to be preserved in non-relativistic limit as well it is natural to keep $T$ the same without scaling. Finally since  an unstable D(p+1)-brane tension is equal to
 $\ttau_{p+1}^{uns}=\frac{\sqrt{2}}{\tl_s^{p+2}}$ we see that it scales as
	\begin{equation}
	\ttau_{p+1}^{uns}=\frac{1}{\omega^{p+2}}\tau_{p+1}^{uns} \ .
	\end{equation}
With the help of these result we find that the matrix 	 $\bA_{\alpha\beta}$ has the form
\begin{eqnarray}
& &\bA_{\alpha\beta}=\omega^2 (\tau_{\alpha\beta}+\partial_\alpha T\partial_\beta T)+\omega (l_s^2 F_{\alpha\beta}+B_{\alpha\beta})+\bh_{\alpha\beta}-
\nonumber \\
& &
-(\tau_\alpha^{ \ A}C_\beta^{ \ B}+\tau_\beta^{ \ A}C_\alpha^{ \ B})\eta_{AB}+
\frac{1}{\omega^2}(m_\alpha^{ \ A}-C_\alpha^{ \ A})(m_\beta^{ \ B}-C_\beta^{ \ B})\eta_{AB} \ .
\nonumber \\
\end{eqnarray}
As the next step we focus on the following expression
\begin{eqnarray}
\tau_{\alpha\beta}+\partial_\alpha T\partial_\beta T=T_\alpha^{ \ \hA}
\eta_{\hA\hB}T_\beta^{ \ \hB} \ , \nonumber \\
\end{eqnarray}
where we introduced object $T_\alpha^{ \ \hA}=(\tau_\alpha^{ \ A},\partial_\alpha T)$
and metric $\eta_{\hA\hB}=\mathrm{diag}(-1,1,\dots,1,1)$.
Now the object $T_\alpha^{ \ \hA}$ is $(p+2)\times (p+2)$ matrix. We presume that   it is  non-singular so that $\bt_{\alpha\beta}=T_\alpha^{ \ \hA}T_\beta^{ \ \hB}\eta_{\hA\hB}$ is non-singular too. Then we can introduce inverse $\tbt^{\alpha\beta}$ that obeys
\begin{equation}
\bt_{\alpha\beta}\tbt^{\beta\gamma}=\delta_\alpha^{ \gamma} \ .
\end{equation}
If we proceed as in previous section we obtain
\begin{eqnarray}
& &\ttau_{p+1}^{uns} e^{-\hphi}\sqrt{-\det \bA_{\alpha\beta}}=\nonumber \\
%=\frac{1}{\omega^{p+2}}\tau_{p+1}^{uns} e^{-\hphi}
%\left(\det (\omega^2 \bt_{\alpha\beta}+
%\bh_{\alpha\beta}-(\tau_\alpha^{ \ A}C_\beta^{ \ B}+\tau_\beta^{ \ A}
%C_\alpha^{ \ B})\eta_{AB}+\frac{1}{\omega^2}(m_\alpha^{ \ A}-C_\alpha^{ \ A})
%(m_\beta^{ \ B}-C_\beta^{ \ B})\eta_{AB}+\right.\nonumber \\
%+\left. \omega (l_s^2F_{\alpha\beta}+B_{\alpha\beta})\right)^{1/2}=\tau_{p+1}^{uns}
%e^{-\hphi}\sqrt{-\det \bt_{\alpha\beta}}\times \nonumber \\
%\times
%\left(\det (\delta^\alpha_\beta+
%\frac{1}{\omega^2}\tbt^{\alpha\gamma}\bh_{\gamma\beta}
%-\frac{1}{\omega^2}\tbt^{\alpha\gamma}(\tau_\gamma^{ \ A}C_\beta^{ \ B}+
%\tau_\beta^{ \ A}C_\gamma^{ \ B})\eta_{AB}+	
%\frac{1}{\omega^4}\tbt^{\alpha\gamma}(m_\gamma^{ \ A}-C_\gamma^{ \ A})
%(m_\beta^{ \ B}-C_\beta^{ \ B})\eta_{AB}\right.+\nonumber \\
%\left.+\frac{1}{\omega} \tbt^{\alpha\gamma}(l_s^2F_{\gamma\beta}+B_{\alpha\beta})\right)^{1/2}	
& &=\tau_{p+1}^{uns} e^{-\hphi}\sqrt{-\det\bt_{\alpha\beta}}\left(1+\frac{1}{2\omega^2}\tbt^{\alpha\beta}\bh_{\alpha\beta}-
\frac{1}{2\omega^2}\tbt^{\alpha\gamma}(\tau_\gamma^{ \ A}C_\alpha^{ \ B}+
\tau_\alpha^{ \ A}C_\gamma^{ \ B})\eta_{AB} -\right.
\nonumber \\
& & \left.-\frac{1}{4\omega^2}\tbt^{\alpha\gamma}(l_s^2F_{\gamma\delta}+B_{\gamma\delta})
\tbt^{\delta\omega}(l_s^2 F_{\omega\alpha}+B_{\omega\alpha})\right) \ .
\nonumber \\
\end{eqnarray}
Since $T_\alpha^{ \ \hA}$ is non-singular it has an inverse matrix
\begin{equation}
T^\alpha_{ \ \hB} \ , \quad T_\alpha^{ \ \hA}T^\beta_{ \ \hA}=\delta_\alpha^\beta \ , \quad
T_\alpha^{ \ \hA}T^\alpha_{ \ \hB}=\delta^{\hA}_{\hB}
\end{equation}
Before we proceed further we perform following extension. We define $X^T_\alpha=\omega\partial_\alpha T-\frac{1}{\omega}C_\alpha^{\  T}$ where
$C_\alpha^{ \ T}=0$. Then we can analyse leading part of WZ term in the following way
\begin{eqnarray}
& &\ttau^{uns}_{p+1}\tl_s dT\wedge \hC^{(p+1)}=\frac{1}{\omega^{p+2}}\tau^{uns}_{p+1} e^{-\hphi}\frac{1}{(p+1)!}\epsilon^{\alpha_1\dots \alpha_{p+2}}\tl_s\partial_{\alpha_1}TX_{\alpha_2}^{ \ A_2}\dots X_{\alpha_{p+2}}^{\ A_{p+2}}
\epsilon_{A_2\dots A_{p+2}}+\nonumber \\
& &+\frac{1}{\omega^{p+1}}\tau_{p+1}^{unst}\omega^{k_{p+1}}l_s dT\wedge C^{(p+1)}= \nonumber \\
%=\frac{1}{\omega^{p+2}}\tau^{uns}_{p+1} e^{-\hphi}\frac{1}{(p+1)!}\epsilon^{\alpha_1\dots \alpha_{p+2}}X^T_{\alpha_1}X_{\alpha_2}^{ \ A_2}\dots X_{\alpha_{p+2}}^{\ A_{p+2}}
%\epsilon_{A_2\dots A_{p+2}}+\nonumber \\
%+\frac{1}{\omega^{p+2}}\tau_{p+1}^{unst}\omega^{k_{p+1}}dT\wedge C^{(p+1)}
%=\nonumber \\
%=\frac{1}{\omega^{p+2}}\tau^{uns}_{p+1} e^{-\hphi}\det X_\alpha^{ \ \hA}
%+\frac{1}{\omega^{p+1}}\tau_{p+1}^{unst}l_s \omega^{k_{p+1}}dT\wedge C^{(p+1)}=
%\nonumber \\
%=\tau_{p+2}^{unst}e^{-\hphi}\det T_\alpha^{ \ \hA}
%\det (\delta_{\hA}^{ \ \hB}-\frac{1}{\omega^2}T^\gamma_{ \ \hB}
%C_\gamma^{ \ \hB})+\frac{1}{\omega^{p+1}}\tau_{p+1}^{unst}l_s \omega^{k_{p+1}}dT\wedge C^{(p+1)}=
%\nonumber \\
& &=\tau_{p+2}^{unst} e^{-\hphi}\det T_\alpha^{ \ \hA}-
\frac{1}{\omega^2}\tau^{unst}_{p+2} e^{-\hphi}T^\alpha_{ \ \hA}
C^{\hA}_{\ \alpha}+
\omega^{k_{p+1}-p-1}\tau^{unst}_{p+1}l_s dT\wedge C^{(p+1)} \ .
\nonumber \\
%--------------
%-----------------------
%\nonumber \\
%=\frac{1}{\omega^{p+1}}\tau_p e^{-\hphi}\det X_\alpha^{ \ A}+\omega^{k_{p+1}-(p+1)}\tau_p C^{(p+1)}=\nonumber \\
%=\tau_p e^{-\hphi}\det  \tau_\alpha^{ \ B}\det(\delta_B^{ \ A}-\frac{1}{\omega^2}\tau_B^{ \ \gamma}C_\gamma^{ \ A})+\omega^{k_{p+1}-(p+1)}\tau_p C^{(p+1)}=\nonumber \\
%=\tau_p e^{-\hphi}\det \tau_\alpha^{ \ A}(1-\frac{1}{\omega^2}\tau_A^{ \ \alpha}C_\alpha^{ \ A})++\omega^{k_{p+1}-(p+1)}\tau_p C^{(p+1)} \ , \nonumber \\
\end{eqnarray}
 %
%
%We see that the kinetic part of the action is singular in the limit $\omega\rightarrow \infty$. This divergence would be cancleled by appropriate form of the $C^{(p+1)}$ form introdced above. Explicitly, we have
%where $X^T_\alpha=\omega\partial_\alpha T-\frac{1}{\omega}C_\alpha^{\  T}$ where
%$C_\alpha^{ \ T}=0$.
We see that two leading order terms in DBI action and WZ action cancel each other. Further,
 in order to have finite WZ term in the action we have to impose condition
\begin{equation}
k_{p+1}=p+1 \ .
\end{equation}
Since  $C_\alpha^{ \ T}=0$  we can write
\begin{equation}
\tau_\alpha^{ \ A}C_\beta^{ \ B}\eta_{AB}=
T_\alpha^{ \ \hA}C_\beta^{ \ \hB}\eta_{\hA\hB}
\end{equation}
so that
\begin{equation}
\tbt^{\alpha\beta}\tau_\alpha^{ \ A}
C_\beta^{ \ B}\eta_{AB}=
\tbt^{\alpha\beta}
T_\beta^{ \ \hA}C_\alpha^{ \ \hB}\eta_{\hA\hB}
=T^\alpha_{ \ \hA}C_\alpha^{ \ \hA}
\end{equation}
and hence we see that all terms proportional to $C_\alpha^{ \ A}$ cancel each other. Further,
in case of the lower dimensional RR fields we find, taking into account expansion of
the exponential function and the fact that $\tl_s^2 \tF_{\alpha\beta}+\tB_{\alpha\beta}=
\omega (l_s^2 F_{\alpha\beta}+B_{\alpha\beta})$ that
\begin{equation}
k_{p-1}=p \ , \quad k_{p-3}=p-1  \ , \dots \ .
\end{equation}
Finally we see that in order to keep finite part of the action we have to have following scaling of dilaton in the form
\begin{equation}
 \hphi=\phi-2\ln \omega
\ .
\end{equation}
%
%Hence $\tbt^{\alpha\beta}=T^\alpha_{ \ \hA}T^\beta_{ \ \hB}\eta^{\hA\hB}$ and
%%hence
%%\begin{equation}
%%\tbt^{\alpha\beta}T_\beta^{ \ \hA}=T^\alpha_{ \ \hC}\eta^{\hC\hA}
%%\end{equation}
%so that we can finally write
%\begin{eqnarray}
%\ttau^{uns}_{p+1} e^{-\hphi}\sqrt{-\det \bA_{\alpha\beta}}=
%\tau_p e^{-\hphi}\det T_\alpha^{ \ \hA}
%+\frac{\tau^{uns}_{p+1}}{2\omega^2}e^{-\hphi}\tau_{p+1} \sqrt{-\det\bt}\tbt^{\alpha\beta}\bh_{\beta\alpha}
%-\frac{\tau^{uns}_{p+1}}{\omega^2}e^{-\hphi}\det T_\alpha^{ \ \hA}
%\tba^{\alpha\beta}\tau_\alpha^{ \ A}C_\beta^{ \ B}\eta_{AB} \ .
%\nonumber \\
%\end{eqnarray}
%
As a result we obtain an action for unstable D(p+1)-brane in the form
\begin{eqnarray}\label{unstaction}
S&=& -\tau^{unst}_{p+1}\int d^{p+2}\xi e^{-\phi}V(T)\sqrt{-\det\bt}
\left[\frac{1}{2}\tbt^{\alpha\beta}\bh_{\alpha\beta}
-\frac{1}{4}\tbt^{\alpha\gamma}(l_s^2F_{\gamma\delta}+B_{\gamma\delta})
\tbt^{\delta\omega}(l_s^2 F_{\omega\alpha}+B_{\omega\alpha})\right]
+\nonumber \\
&+&\tau_{p+1}^{unst} \int d^{p+2}\xi l_s  V(T)dT\wedge  (C^{(p+1)}+C^{(p-1)}\wedge (l_s^2 F+B)+\dots) \ .
\nonumber \\
\end{eqnarray}
We would like to stress a crucial difference between this action and the
action studied in \cite{Roychowdhury:2019qmp}. The crucial fact is the action
(\ref{unstaction})
depends on the derivative of tachyon through the matric $\bt$ and that even
if the action is $p+2$ dimensional we work with $\tau_\mu^{ \ A}$ where $A=0,1,\dots,p$. This
is natural choice since $T$ can be considered as an additional coordinate, see for
example \cite{Erkal:2009xq}.
As a check of our proposal we will study tachyon kink solution in the next section.

\section{Stable Dp-Brane as Gauge Fixed Unstable D(p+1)-Brane}\label{fourth}
We are going to argue that tachyon kink solution corresponds to stable non-relativistic
Dp-brane. To do this we will follow \cite{Erkal:2009xq} and interpret stable
Dp-brane as gauge fixing unstable D(p+1)-brane that is stretched along one spatial
dimension, $\xi^{p+1}\equiv y$. In this case we can presume that $T=f(y)$ and all
remaining fields do not depend on $y$ as well. We further partially fix the gauge
and impose condition $A_y=0$. Let us denote remaining world-volume coordinates
as $\xi^{\halpha} \ , \ \halpha,\hbeta=0,1,\dots,p$. Then the matrix $\bt_{\alpha\beta}$ has the form
\begin{equation}
\bt_{\halpha\hbeta}=\hbt_{\halpha\hbeta} \ , \quad \hbt_{\halpha\hbeta}=\tau_{\halpha\hbeta} \ , \quad
\bt_{\halpha y}=0 \ ,  \quad \bt_{yy}=l_s^2f'^2 \ ,
\end{equation}
where we presume $T=f(y)$. Then clearly
\begin{equation}
\det \bt=l_s^2f'^2\det \hbt \ , \quad \tbt^{\alpha\beta}=
\left(\begin{array}{cc}
\hbt^{\halpha\hbeta} & 0 \\
0 & \frac{1}{f'^2} \\ \end{array}\right)  \ .
\end{equation}
Inserting this ansatz into the action  (\ref{unstaction}) and using the fact that
\begin{equation}
\bh_{y\alpha}=0 \ , \quad
F_{y\alpha}=0
\end{equation}
we get
\begin{eqnarray}\label{gaugefixed}
& &S=-\tau_{p+1}^{unst}l_s
\int dy f' V(f) \int d^{p+1}\xi e^{-\phi}
\sqrt{-\det \ba}\left[
\frac{1}{2}\tba^{\halpha\hbeta}\bh_{\halpha\hbeta}-\right.\nonumber \\
& &-
\left.\frac{1}{4}\tba^{\halpha\hgamma}(l_s^2 F_{\hgamma\hdelta}+B_{\hgamma\hdelta})
\tba^{\hdelta\homega}(l_s^2 F_{\homega\halpha}+B_{\homega\halpha})\right]+\nonumber \\
& &+\tau_{p+1}^{unst}l_s \int dy f'V(f)\int d^{p+1}\xi   (C^{(p+1)}+C^{(p-1)}\wedge (l_s^2 F+B)+\dots) \nonumber \\
\end{eqnarray}
so that when we identify
\begin{equation}
\tau_p=\tau^{unst}_{p+1}l_s\int dy f'(y) V(f)
\end{equation}
we see that (\ref{gaugefixed}) corresponds to the action of non-relativistic
Dp-brane action that was introduced in previous section.
\section{Hamiltonian Formalism}\label{fifth}
We showed in the previous section that the tachyon kink solution on the world-volume of unstable
non-relativistic D(p+1)-brane corresponds to stable non-relativistic Dp-brane. However there is a question of the fate of this unstable object when the tachyon is in its minimum everywhere on its worldvolume. We know that in case of relativistic unstable Dp-brane the tachyon vacuum corresponds to the gas of fundamental strings with agreement with the basic picture of the tachyon condensation
\cite{Sen:2004nf,Sen:2000kd,Sen:2003bc}. It is natural to ask whether similar behaviour occurs in case of non-relativistic D(p+1)-brane as well. In order to answer it we shall find Hamiltonian for this unstable D(p+1)-brane.

 Recall
that the action for non-relativistic unstable D(p+1)-brane  has the form
\begin{eqnarray}
S=-\tau^{unst}_{p+1}\int d^{p+2}\xi e^{-\phi}V(T)\sqrt{-\det\bt}
\left[\frac{1}{2}\tbt^{\alpha\beta}\bh_{\alpha\beta}
-\frac{1}{4}\tbt^{\alpha\gamma}(l_s^2F_{\gamma\delta}+B_{\gamma\delta})
\tbt^{\delta\omega}(l_s^2 F_{\omega\alpha}+B_{\omega\alpha})\right]  \ , \nonumber \\
\end{eqnarray}
where we restrict ourselves to the case of vanishing RR field.
To proceed further let us elaborate following point. We have non-singular
matrix $T_\alpha^{ \ \hA}$ so that
\begin{equation}
\bt_{\alpha\beta}=T_\alpha^{ \ \hA}T_\beta^{ \ \hB}\eta_{\hA\hB} \ ,
\end{equation}
we can proceed further and introduce $x^{M}=(x^\mu,T)$ so that
\begin{equation}
T_\alpha^{ \ \hA}=\partial_\alpha x^M T_M^{ \ \hA} \ .
\end{equation}
We also extend $\bh_{\mu\nu}$ as $\bh_{MN}=(\bh_{\mu\nu},0)$ so that
\begin{equation}
\bh_{\alpha\beta}=\partial_\alpha x^M\bh_{MN}\partial_\beta x^N \ .
\end{equation}
Then
\begin{equation}
\sqrt{-\det\bt}=\det T_\alpha^{ \ \hA} \  , \quad \tbt^{\alpha\beta}=T^{\alpha }_{ \ \hA}
T^{\beta}_{ \ \hB}\eta^{\hA\hB} \ .
\end{equation}
Now
\begin{eqnarray}
& &\frac{\delta \bt_{\alpha\beta}}{\delta \partial_0 x^M}=\delta_\alpha^0 T_M^{ \ \hA}
T_\beta^{ \ \hB}\eta_{\hA\hB}+T_\alpha^{ \ \hA}\eta_{\hA\hB}T_M^{ \ \hB}\delta^0_\beta
\ , \nonumber \\
& &\frac{\delta \tbt^{\alpha\beta}}{\delta \partial_0 x^M}
%=-
%\tbt^{\alpha\gamma}\frac{\delta \bt_{\gamma\delta}}{\delta \partial_0 x^M}
%\tbt^{\delta\beta}
=-\bt^{\alpha\gamma}T_\gamma^{ \ \hA}
\eta_{\hA\hB}T_M^{ \ \hB}\bt^{0\beta}
-\bt^{\alpha 0}T_M^{ \ \hA}T_\delta^{ \ \hB}\eta_{\hA\hB}\bt^{\delta \beta}
\nonumber \\
& &\frac{\delta \det  T_\alpha^{ \ \hA}}{\delta \partial_0 x^M}=T_M^{ \ \hA}
T^0_{ \ \hA} \det T_\alpha^{ \ \hA} \ . \nonumber \\
\end{eqnarray}
Then we have following conjugate momenta
\begin{eqnarray}
& &p_M=\frac{\delta L}{\delta \partial_0 x^M}=
-\tau_{p+1}^{unst} e^{-\phi}V
T_M^{ \ \hA}T^0_{ \ \hA}\det T_\alpha^{ \ \hA}
\bX+\nonumber \\
& &+\tau_{p+1}^{unst}e^{-\phi}V
\det T_\alpha^{\ \hA}[\tbt^{0\beta}T_M^{ \ \hB}\eta_{\hB\hA}
T_\gamma^{  \ \hA}\tbt^{\gamma\alpha}\bh_{\alpha\beta}+
\bh_{MN}\partial_\alpha x^N\tbt^{\alpha 0}-\nonumber \\
& &-T_M^{ \ \hA}\eta_{\hA\hB}T_{\gamma'}^{ \ \hB}
\tbt^{\gamma'\alpha}\mF_{\alpha\omega}
\tbt^{\omega\delta}\mF_{\delta\gamma}\tbt^{\gamma 0}+
B_{MN}\partial_\delta x^N \tbt^{\delta\omega}\mF_{\omega\alpha}\tbt^{\alpha 0}] \ ,
\nonumber \\
\end{eqnarray}
where
\begin{equation}
\bX=
\frac{1}{2}\tbt^{\alpha\beta}\bh_{\alpha\beta}
-\frac{1}{4}\tbt^{\alpha\gamma}\mF_{\gamma\delta}
\tbt^{\delta\omega}\mF_{\omega\alpha} \ , \quad \mF_{\alpha\beta}=l_s^2 F_{\alpha\beta}+B_{\alpha\beta} \ .
\end{equation}
Further, momentum conjugate to $A_\alpha$ is equal to
\begin{eqnarray}
\pi^\alpha=\frac{\delta L}{\delta \partial_0 A_\alpha}=\tau_{p+1}^{unst}
l_s^2 e^{-\phi}V\det T_\alpha^{ \ \hA}\tbt^{\alpha\gamma}\mF_{\gamma\delta}\tbt^{\delta 0} \ .  \nonumber \\
\end{eqnarray}
 Now since
$\mF_{\alpha\beta}=-\mF_{\beta\alpha}$ we immediately obtain primary constraint $\pi^0\approx 0$.
Then with the help of  $\pi^i$ defined above  we can introduce $\Pi_M$ as
\begin{eqnarray}
\Pi_M=p_M-
l_s^{-2}B_{MN}\partial_i x^N \pi^i=
-\tau_{p+1}^{unst} e^{-\phi}V
T_M^{ \ \hA}T^0_{ \ \hA}\det T_\alpha^{ \ \hA}
\bX+\nonumber \\
+\tau_{p+1}^{unst}e^{-\phi}V
\det T_\alpha^{\ \hA}[\bt^{0\beta}T_\beta^{ \ \hB}\eta_{\hB\hA}
T_\gamma^{  \ \hA}\bt^{\gamma\alpha}\bh_{\alpha\beta}+
\bh_{MN}\partial_\alpha x^N\bt^{\alpha 0}] \ . \nonumber \\
\nonumber \\
\end{eqnarray}
%using the fact htat
%\begin{eqnarray}
%-\tau_{p+1}^{unst}e^{-\phi}V\det T_\alpha^{ \ \hA}T_M^{ \ \hA}\eta_{\hA\hB}T_{\gamma'}^{ \ \hB}
%\tbt^{\gamma'\alpha}(l_s^2 F_{\alpha\omega}+B_{\alpha\omega})
%\tbt^{\omega\delta}(l_s^2 F_{\delta\gamma}+B_{\delta\gamma})\tbt^{\gamma 0}=
%\nonumber \\
%=-\tau_{p+1}^{unst}e^{-\phi}V\det T_\alpha^{ \ \hA}T_M^{ \ \hA}\eta_{\hA\hB}T_{\gamma'}^{ \ \hB}
%\tbt^{\gamma'\alpha}(l_s^2 F_{\alpha \omega}+B_{\alpha \omega})
%\tbt^{\omega\delta'}\bt_{\delta'\alpha'}\tbt^{\alpha'\delta}(l_s^2 F_{\delta\gamma}+B_{\delta\gamma})\tbt^{\gamma 0}=
%\nonumber \\
%=-\det T_\alpha^{ \ \hA}T_M^{ \ \hA}\eta_{\hA\hB}T_{\gamma'}^{\ \hB}
%\tbt^{\gamma'\alpha}(l_s^2 F_{\alpha \omega}+B_{\alpha \omega})
%\tbt^{\omega\delta'}\bt_{\delta' i}\pi^i \nonumber \\
%\end{eqnarray}
Now we have to introduce metric $h^{MN}$. Since
\begin{equation}
T_M^{ \ \hA}=\left(\begin{array}{cc}
\tau_\mu^{ \ A} & 0 \\
0 & 1 \\ \end{array}\right)
\end{equation}
we see that  it is natural to choose $h^{MN}$ as
\begin{equation}
h^{MN}=\left(\begin{array}{cc}
h^{\mu\nu} & 0 \\
0 & 0 \\ \end{array}\right)\equiv E^M_{ \ A'}E^N_{ \ B'}\delta^{A'B'}
\end{equation}
and also $T^M_{ \ \hB}$ as
\begin{equation}
T^M_{ \ \hB}=\left(\begin{array}{cc}
\tau^\mu_{ \ A} & 0 \\
0 & 1 \\ \end{array} \right)
\ , \quad  T^M_{ \ \hA}T_M^{ \ \hB}=\delta_{\hA}^{\hB}
\end{equation}
that obeys the relation
\begin{equation}
h^{MN}T_N^{ \ \hA}=0 \ .
\end{equation}
Let us introduce following object
\begin{equation}
\tE_M^{ \ A'}=E_M^{ \ A'}-m_N^{ \ \hB}E^N_{ \ C'}\delta^{C'A'}\tau_M^{ \ \hA}\eta_{\hA\hB} \ ,
\quad
\tT^M_{ \ \hA}=T^M_{ \ \hA}-h^{MN}m_N^{\ \hB}\eta_{\hB\hA}
\end{equation}
that clearly obeys
\begin{equation}
\tT^M_{ \ A}\tE_M^{ \ A'}=0 \ , \quad \tE_M^{ \ A'}\delta_{A'B'}\tE_N^{ \ B'}=
\bh_{MN}+\tau_M^{ \ \hA}\Phi_{\hA\hB}\tau_N^{ \ \hB} \ , \quad  \tT^M_{ \ \hA}T_M^{ \ \hB}=0 \ ,
\end{equation}
where
\begin{equation}
\Phi_{\hA\hB}=-T^M_{ \ \hA}m_M^{ \ \hC}\eta_{\hC\hB}-
\eta_{\hA\hC}m_M^{ \ \hC}T^M_{ \ \hB}+
\eta_{\hA\hC}
m_M^{ \ \hC}h^{MN}m_N^{ \ \hD}\eta_{\hD\hB} \ .
\end{equation}
Then we obtain
\begin{eqnarray}
\Pi_M h^{MN}\Pi_N=(\tau_{p+1}^{unst}e^{-\phi}V \det T_\alpha^{ \ \hA})^2
\tbt^{0\alpha}\partial_\alpha x^N \tE_N^{ \ A'}\delta_{A' B' }\tE_M^{ \ B'}\partial_\beta x^M
\tbt^{\beta 0}\  \nonumber \\
\end{eqnarray}
and also
\begin{eqnarray}
& &\Pi_M\tT^M_{ \ \hA}=-\tau_{p+1}^{unst}e^{-\phi}V \det T_\alpha^{ \ \hA}T^0_{ \ \hA}\bX
+\tau^{unst}_{p+1}e^{-\phi}V\det T_\alpha^{ \ \hA}\tbt^{0\beta}\eta_{\hA\hB}T_\gamma^{ \ \hB}
\tbt^{\gamma\alpha}\bh_{\alpha\beta}-\nonumber \\
& &-\tau_{p+1}^{unst}e^{-\phi}V\det T_\alpha^{ \ \hA}\Phi_{\hA\hB}\tau_N^{ \ \hB}\partial_\alpha x^N \tbt^{\alpha 0}
-\tau^{unst}_{p+1}e^{-\phi}V \det T_\alpha^{ \ \hA}
\eta_{\hA\hB}T_\beta^{ \ \hB}
\tbt^{\beta\alpha}\mF_{\alpha\omega}\tbt^{\omega\delta}\mF_{\delta\gamma}\tbt^{\gamma 0} \ .
\nonumber \\
\end{eqnarray}
Let us multiply this result with the expression
\begin{equation}
\frac{\tau_{p+1}^{unst}e^{-\phi}V}{(p+1)!}
\eta^{\hA\hA_0}\epsilon_{\hA_0\hA_1\dots \hA_{p+1}}
T_{i_1}^{\hA_1}\dots T_{i_{p+1}}^{\hA_{p+1}}
\epsilon^{i_1\dots i_{p+1}} \ .
\end{equation}
Then, after some tedious algebra, we obtain
\begin{eqnarray}
&&\Pi_M T^M_{ \ \hA}
\frac{\tau_{p+1}^{unst}e^{-\phi}V}{(p+1)!}
\eta^{\hA\hA_0}\epsilon_{\hA_0\hA_1\dots \hA_{p+1}}
T_{i_1}^{\hA_1}\dots T_{i_{p+1}}^{\hA_{p+1}}
\epsilon^{i_1\dots i_{p+1}}-\Pi_M h^{MN}\Pi_N=\nonumber \\
& &=
-(\tau^{unst}_{p+1} e^{-\phi}V^2)
(\det \hbt_{ij} \bX
+ \det\hbt	\tbt^{0\alpha}\mF_{\alpha\omega}\tbt^{\omega\delta}
	\mF_{\delta \gamma}\tbt^{\gamma 0}) \ .
	\nonumber \\
	\end{eqnarray}
To proceed further  we have to introduce explicit form of inverse matrix $\tbt^{\alpha\beta}$
\begin{eqnarray}
& &\tbt^{00}=\frac{\det \hbt_{ij}}{\det \bt} \ , \quad \tbt^{0i}=-\bt_{0k}\hbt^{kj}\frac{\det\hbt_{ij}}{\det \bt} \ , \nonumber \\
& &\tbt^{i0}=-\hbt^{ik}\bt_{k0}\frac{\det \hbt_{ij}}{\det \bt} \ , \quad
\tbt^{ij}=\hbt^{ij}+\frac{\det \hbt_{ij}}{\det \bt}
\hbt^{ik}\bt_{k0}\bt_{0l}\hbt^{lj} \ , \nonumber \\
\end{eqnarray}
where $ \hbt^{ij}$ is $(p+1)\times (p+1)$ matrix inverse to $\hbt_{ij}$ so that
\begin{equation}
	\hbt^{ij}\hbt_{jk}=\delta^i_j \ .
	\end{equation}
With the help of these formulas we obtain	
%\begin{eqnarray}
%& &\Pi_M h^{MN}\Pi_N=-(\tau_{p+1}^{nst}e^{-\phi}V)^2	\det\bt (\tbt^{00}\tE_0^{ \ A'}
%\tE_0^{ \ B'}\delta_{A'B'}\tbt^{00}+2\tbt^{00}\tbt^{0i}
%\tE_0^{ \ A'}\tE_i^{ \ B'}\delta_{A' B'}+\tbt^{0i}\tbt^{0j}\tE_i^{ \ A'}\tE_j^{ \ A'}
%\delta_{A' B'})=\nonumber \\
%=-(\tau_{p+1}^{unst}e^{-\phi}V)\det \hbt_{ij}
%(\tE_0^{ \ A'}
%\tE_0^{ \ B'}\delta_{A'B'}\tbt^{00}+2\tbt^{0i}
%\tE_0^{ \ A'}\tE_i^{ \ B'}\delta_{A' B'}+\frac{\tbt^{0i}\tbt^{0j}}{\tbt^{00}}
%\tE_i^{ \ A'}\tE_j^{ \ B'}\delta_{A' B' })
%\nonumber \\
%\Rightarrow
%-\frac{1}{(\tau_{p+1}^{unst}e^{-\phi}V)^2\det \hbt_{ij}}\Pi_Mh^{MN}\Pi_N
%-\frac{\tbt^{0i}\tbt^{0j}}{\tbt^{00}}
%\tE_i^{ \ A'}\tE_j^{ \ B'}\delta_{A' B'}=
%\tE_0^{ \ A'}
%\tE_0^{ \ B'}\delta_{A'B'}\tbt^{00}+2\tbt^{0i}
%\tE_0^{ \ A'}\tE_i^{ \ B'}\delta_{A' B'} \nonumber \\
%\end{eqnarray}
%and we use the fact that
%\begin{eqnarray}
%\tbt^{\alpha\beta}\bh_{\beta\alpha}=\tbt^{\alpha\beta}
%(\tE_\alpha^{ \ A'}\tE_\beta^{ \ B'}
%-T_\alpha^{ \ \hA}\Phi_{\hA\hB}T_\beta^{ \ \hB})=\nonumber \\
%\tbt^{\alpha\beta}\tE_\alpha^{ \ A'}\tE_\beta^{ \ B'}\delta_{A' B' }-\eta^{AB}\Phi_{AB}= \nonumber \\
%=-\frac{1}{(\tau_{p+1}^{unst}e^{-\phi}V)^2\det \hbt_{ij}}\Pi_Mh^{MN}\Pi_N+
%\hbt^{ij}\tE_i^{ \ A'}\tE_j^{ \ B'}\delta_{A' B' }-
%\eta^{AB}\Phi_{AB} \nonumber \\
%\nonumber \\
%\end{eqnarray}	
%and hence we obtain
\begin{eqnarray}
& &\Pi_M T^M_{ \ \hA}
\frac{\tau_{p+1}^{unst}e^{-\phi}V}{(p+1)!}
\eta^{\hA\hA_0}\epsilon_{\hA_0\hA_1\dots \hA_{p+1}}
T_{i_1}^{\hA_1}\dots T_{i_{p+1}}^{\hA_{p+1}}
\epsilon^{i_1\dots i_{p+1}}-\frac{1}{2}\Pi_M h^{MN}\Pi_N+\nonumber \\
& &\frac{1}{2}(\tau^{unst}_{p+1}e^{-\phi }V)^2\det \hbt \hbt^{ij}
\tE_i^{ \ A'}\tE_j^{ \ B'}\delta_{A'B'}-\frac{1}{2}\det\hbt
\eta^{AB}\Phi_{AB}
=\nonumber \\
& &-(\tau_{p+1}^{unst}e^{-\phi}V)^2(-\frac{1}{4}\det \hbt
\tbt^{\alpha\gamma}
\mF_{\gamma\omega}\tbt^{\omega\delta}\mF_{\delta\alpha}
+\det \bt \tbt^{0\alpha}\mF_{\alpha\omega}\tbt^{\omega\delta}\mF_{\delta\gamma}\tbt^{\gamma 0}) \ .
\nonumber \\
\end{eqnarray}
As the last step we use the fact that
\begin{equation}
F_{\alpha\delta}\tbt^{\delta 0}=\frac{1}{\tau_{p+1}^{unst}l_s^2 e^{-\phi}V}
\bt_{\alpha\gamma}\pi^\gamma
\end{equation}
so that
%\begin{equation}
%\det \bt \tbt^{0\alpha}\mF_{\alpha\omega}\tbt^{\omega\delta}\mF_{\delta\gamma}\tbt^{\gamma 0}
%=\frac{1}{(\tau_{p+1}^{unst}l_s^2 e^{-\phi}V)^2}\pi^i\bt_{ij}\pi^j
%\end{equation}	
%To proceed further we have to perform following manipulation
\begin{eqnarray}
\tbt^{\alpha\gamma}	\mF_{\gamma\delta}
\tbt^{\delta \omega}\mF_{\omega\alpha}=
%=2\tbt^{00}\mF_{0j}\tbt^{ji}\mF_{j0}+
%2\tbt^{0i}\mF_{i0}\tbt^{0j}\mF_{j0}+\nonumber \\
%+4\tbt^{0i}\mF_{ij}\tbt^{ji}\mF_{i0}+\tbt^{ij}\mF_{jk}
%\tbt^{kl}\mF_{li}=\nonumber \\
%=-2\tbt^{00}(\mF_{i0}+\frac{1}{\tbt^{00}}\mF_{ik}\tbt^{k0})\hbt^{ij}
%(\mF_{j0}+\frac{1}{\tbt^{00}}\mF_{jl}\tbt^{l0})+\hbt^{ij}\mF_{jk}\hbt^{kl}\mF_{li}=\nonumber \\
\frac{2}{(\tau_{p+1}^{unst}e^{-\phi}V )^2\det \tbt}\pi^i\tbt_{ij}\pi^j+
\hbt^{ij}\mF_{jk}\hbt^{kl}\mF_{li} \ .
\nonumber \\
\end{eqnarray}
%	using the fact that
%\begin{equation}
%\pi^i=(\tau_{p+1}^{unst}l_s^2e^{-\phi}V\det T)\tbt^{00}(\hbt^{ij}\mF_{j0}+\frac{1}{\tbt^{00}}\tbt^{ij}\mF_{jk}\tbt^{k0})	
%=	\tbt^{00}(\tau_{p+1}^{unst}l_s^2e^{-\phi}V\det T)\hbt^{ij}(\mF_{j0}+\frac{1}{\tbt^{00}}\mF_{jk}\tbt^{k0})
%\end{equation}
Then collecting there terms together we obtain desired result which
is the Hamiltonian constraint
\begin{eqnarray}
& &\mH_\tau=
-\Pi_M T^M_{ \ \hA}
\frac{\tau_{p+1}^{unst}e^{-\phi}V}{(p+1)!}
\eta^{\hA\hA_0}\epsilon_{\hA_0\hA_1\dots \hA_{p+1}}
T_{i_1}^{\hA_1}\dots T_{i_{p+1}}^{\hA_{p+1}}
\epsilon^{i_1\dots i_{p+1}}-\frac{1}{2}\Pi_M h^{MN}\Pi_N+\frac{1}{2}\pi^i\hbt_{ij}\pi^j+\nonumber \\
& &+\frac{1}{2}(\tau_{p+1}e^{-\phi} V)^2\det \hbt [\hbt^{ij}
\tE_i^{ \ A'}\tE_j^{ \ B'}\delta_{A'B'}+
\eta^{AB}\Phi_{AB}
+\frac{1}{2}\hbt^{ij}\mF_{jk}\hbt^{kl}\mF_{li}]
\approx  0 \ . \nonumber \\
\end{eqnarray}
It is clear that it case of stable D(p+1)-brane the Hamiltonian constraint has the same form
when replace $\tau^{unst}_{p+1}$ with $\tau_{p+1}$ and when we impose  $V(T)=1$. Alternatively, we can get Hamiltonian for stable Dp-brane by gauge fixing as was performed in section (\ref{third}) in Lagrangian formalism. However we are mainly interested in the tachyon vacuum
condensation that corresponds to configuration
%With the help of this constraint we can study tachyon condensation using the approach
%when the tachyon condensation corresponds to the gauge fixing of one world-volume coordinate
%\cite{Erkal:2009xq}. Explicitly, we fix the gauge as
%\begin{equation}
%T=f(a(y)) \ , f'>0 \ , \xi^{p+1}\equiv y  ,  A_y=0
%\end{equation}
%and all remaining fields depend on $\xi^{\hi} \ , \hi=1,\dots,p$ only. Then the spatial diffeomorphism constraint $\mH_y\approx 0$ gives $p_T=0$ since $A_y=0$. Then we have $\Pi_T=0$
%and hence Hamiltonian constraint has the form
%\begin{eqnarray}
%\mH_{\tau}^{unst}=Vaf'(ay)\Pi_\mu \tau^\mu_{ \ A}e^{-\phi}\eta^{AA_0}\epsilon_{A_0\dots A_{p}}
%T_{\hi_1}^{ \ A_1}\dots T_{\hi_p}^{ \ A_p}-\Pi_\mu h^{\mu\nu}\Pi_\nu
%-
%\pi^{\hi}\bt_{\hi \hj}\pi^{\hj}+\nonumber \\
%+
%\frac{1}{2}(\tau_{p+1}V)^2e^{-2\phi}af'(ay)\det \bt_{\hi\hj}\tbt^{\hi\hj}
%\hE_{\hi}^{ \ A'}E_{\hj}^{ \ B'}\delta_{A'B'}
%-\frac{1}{2}(\tau_{p+1}V)^2af'(y)\det \bt
%\eta^{AB}\Phi_{AB}-\frac{1}{2}(\tau_{p+1}^{unst}V)^2af'(ay)e^{-2\phi}
%\tbt^{\hi\hj}\mF_{jk}\tbt^{\hk\hl}\mF_{\hl\hi}
%\nonumber \\
%\end{eqnarray}
\begin{equation}
T=T_{\min} \ , \quad  V(T_{min})=0 \ .
\end{equation}
In order to see the fate of the tachyon vacuum state let us insert these values into the Hamiltonian constraint given above.
 In this case the Hamiltonian constraint takes the form
\begin{equation}
\mH_\tau(T=T_{min})=\frac{1}{2}\Pi_\mu h^{\mu\nu}\Pi_\nu+\pi^i\hbt_{ij}\pi^j
 \approx 0 \ .
\end{equation}	
We see that this result does not correspond to either relativistic or non-relativistic string.
This is rather unsatisfactory result especially when we compare with the vacuum tachyon condensation in case of relativistic unstable D-brane \cite{Sen:2000kd,Sen:2003bc} where it was shown that the tachyon vacuum state corresponds to the gas of fundamental strings whose equations of motion can be easily derived from the Hamiltonian constraint for unstable Dp-brane evaluated at the tachyon vacuum. Recall that similar situation occurs in case of the unstable Dp-brane in torsional NC background as was recently shown in
\cite{Kluson:2020aoq}. We mean that this unsatisfactory result with the tachyon condensation on the world-volume of unstable D(p+1)-brane in p-brane NC background has its origin in the way we define these objects when we cancel divergence in DBI part of the action with an appropriate term in WZ term while non-relativistic string is defined by similar procedure however with an appropriate NSNS two form.  For that reason we expect that satisfactory result of the vacuum tachyon condensation on the world-volume of non-relativistic D(p+1)-brane could be achieved when we consider more general form of the limiting procedure with non-trivial RR and NSNS two forms. This situation is currently under study.


\begin{thebibliography}{20}
	
%\cite{Cartan:1923zea}
\bibitem{Cartan:1923zea}
E.~Cartan,
\emph{``Sur les variétés à connexion affine et la théorie de la relativité généralisée. (première partie),''}
Annales Sci.\ Ecole Norm.\ Sup.\  {\bf 40} (1923) 325.
%%CITATION = ASENA,40,325;%%
%147 citations counted in INSPIRE as of 02 Nov 2017



	
%\cite{Christensen:2013lma}
\bibitem{Christensen:2013lma}
M.~H.~Christensen, J.~Hartong, N.~A.~Obers and B.~Rollier,
\emph{``Torsional Newton-Cartan Geometry
	and Lifshitz Holography,''}
Phys.\ Rev.\ D {\bf 89} (2014) 061901
doi:10.1103/PhysRevD.89.061901
[arXiv:1311.4794 [hep-th]].
%%CITATION = doi:10.1103/PhysRevD.89.061901;%%
%77 citations counted in INSPIRE as of 15 Jan 2018	

%\cite{Christensen:2013rfa}
\bibitem{Christensen:2013rfa}
M.~H.~Christensen, J.~Hartong, N.~A.~Obers and B.~Rollier,
\emph{``Boundary Stress-Energy Tensor and Newton-Cartan Geometry in Lifshitz Holography,''}
JHEP {\bf 1401} (2014) 057
doi:10.1007/JHEP01(2014)057
[arXiv:1311.6471 [hep-th]].
%%CITATION = doi:10.1007/JHEP01(2014)057;%%
%78 citations counted in INSPIRE as of 15 Jan 2018
	
	%\cite{Hartong:2014oma}
	\bibitem{Hartong:2014oma}
	J.~Hartong, E.~Kiritsis and N.~A.~Obers,
\emph{"Lifshitz space–times for Schrödinger holography,''}
	Phys.\ Lett.\ B {\bf 746} (2015) 318
	doi:10.1016/j.physletb.2015.05.010
	[arXiv:1409.1519 [hep-th]].
	%%CITATION = doi:10.1016/j.physletb.2015.05.010;%%
	%65 citations counted in INSPIRE as of 15 Jan 2018




%\cite{Hansen:2020pqs}
\bibitem{Hansen:2020pqs}
D.~Hansen, J.~Hartong and N.~A.~Obers,
\emph{``Non-Relativistic Gravity and its Coupling to Matter,''}
arXiv:2001.10277 [gr-qc].
%%CITATION = ARXIV:2001.10277;%%
%2 citations counted in INSPIRE as of 20 Mar 2020

%\cite{Hansen:2019svu}
\bibitem{Hansen:2019svu}
D.~Hansen, J.~Hartong and N.~A.~Obers,
\emph{``Non-relativistic expansion of the Einstein-Hilbert Lagrangian,''}
arXiv:1905.13723 [gr-qc].
%%CITATION = ARXIV:1905.13723;%%
%7 citations counted in INSPIRE as of 20 Mar 2020

%\cite{Harmark:2017rpg}
\bibitem{Harmark:2017rpg}
T.~Harmark, J.~Hartong and N.~A.~Obers,
\emph{``Nonrelativistic
	strings and limits of the AdS/CFT correspondence,''}
Phys.\ Rev.\ D {\bf 96} (2017) no.8,  086019
doi:10.1103/PhysRevD.96.086019
[arXiv:1705.03535 [hep-th]].
%%CITATION = doi:10.1103/PhysRevD.96.086019;%%
%5 citations counted in INSPIRE as of 15 Jan 2018







%\cite{Kluson:2020aoq}
\bibitem{Kluson:2020aoq}
J.~Klusoň,
\emph{``Unstable D-brane in Torsional Newton-Cartan Background,''}
arXiv:2001.11543 [hep-th].
%%CITATION = ARXIV:2001.11543;%%
%1 citations counted in INSPIRE as of 21 Mar 2020


%\cite{Roychowdhury:2020kma}
\bibitem{Roychowdhury:2020kma}
D.~Roychowdhury,
\emph{``Nonrelativistic giant magnons from Newton Cartan strings,''}
JHEP {\bf 2002} (2020) 109
doi:10.1007/JHEP02(2020)109
[arXiv:2001.01061 [hep-th]].
%%CITATION = doi:10.1007/JHEP02(2020)109;%%
%1 citations counted in INSPIRE as of 21 Mar 2020

%\cite{Kluson:2019ajy}
\bibitem{Kluson:2019ajy}
J.~Klusoň,
\emph{``Canonical Description of T-duality with NSNS Background,''}
arXiv:1911.13001 [hep-th].
%%CITATION = ARXIV:1911.13001;%%

%\cite{Pereniguez:2019eoq}
\bibitem{Pereniguez:2019eoq}
D.~Pereñiguez,
\emph{``$p$-brane Newton–Cartan geometry,''}
J.\ Math.\ Phys.\  {\bf 60} (2019) no.11,  112501
doi:10.1063/1.5126184
[arXiv:1908.04801 [hep-th]].
%%CITATION = doi:10.1063/1.5126184;%%
%3 citations counted in INSPIRE as of 21 Mar 2020

%\cite{Kluson:2019avy}
\bibitem{Kluson:2019avy}
J.~Klusoň,
\emph{``Non-Relativistic D-brane from T-duality Along Null Direction,''}
JHEP {\bf 1910} (2019) 153
doi:10.1007/JHEP10(2019)153
[arXiv:1907.05662 [hep-th]].
%%CITATION = doi:10.1007/JHEP10(2019)153;%%
%4 citations counted in INSPIRE as of 21 Mar 2020


%\cite{Harmark:2019upf}
\bibitem{Harmark:2019upf}
T.~Harmark, J.~Hartong, L.~Menculini, N.~A.~Obers and G.~Oling,
\emph{``Relating non-relativistic string theories,''}
JHEP {\bf 1911} (2019) 071
doi:10.1007/JHEP11(2019)071
[arXiv:1907.01663 [hep-th]].
%%CITATION = doi:10.1007/JHEP11(2019)071;%%
%20 citations counted in INSPIRE as of 21 Mar 2020

%\cite{Harmark:2018cdl}
\bibitem{Harmark:2018cdl}
T.~Harmark, J.~Hartong, L.~Menculini, N.~A.~Obers and Z.~Yan,
\emph{``Strings with Non-Relativistic Conformal Symmetry and Limits of the AdS/CFT Correspondence,''}
JHEP {\bf 1811} (2018) 190
doi:10.1007/JHEP11(2018)190
[arXiv:1810.05560 [hep-th]].
%%CITATION = doi:10.1007/JHEP11(2018)190;%%
%35 citations counted in INSPIRE as of 21 Mar 2020



%\cite{Kluson:2018egd}
\bibitem{Kluson:2018egd}
J.~Klusoň,
\emph{``Remark About Non-Relativistic
	 String in Newton-Cartan Background and Null Reduction,''}
JHEP {\bf 1805} (2018) 041
doi:10.1007/JHEP05(2018)041
[arXiv:1803.07336 [hep-th]].
%%CITATION = doi:10.1007/JHEP05(2018)041;%%
%21 citations counted in INSPIRE as of 21 Mar 2020



%\cite{Andringa:2012uz}
\bibitem{Andringa:2012uz}
R.~Andringa, E.~Bergshoeff, J.~Gomis and M.~de Roo,
\emph{``'Stringy' Newton-Cartan Gravity,''}
Class.\ Quant.\ Grav.\  {\bf 29} (2012) 235020
doi:10.1088/0264-9381/29/23/235020
[arXiv:1206.5176 [hep-th]].
%%CITATION = doi:10.1088/0264-9381/29/23/235020;%%
%14 citations counted in INSPIRE as of 06 Nov 2017





%\cite{Kluson:2018uss}
\bibitem{Kluson:2018uss}
J.~Klusoň,
\emph{"Hamiltonian for a string in a Newton-Cartan background,''}
Phys.\ Rev.\ D {\bf 98} (2018) no.8,  086010
doi:10.1103/PhysRevD.98.086010
[arXiv:1801.10376 [hep-th]].
%%CITATION = doi:10.1103/PhysRevD.98.086010;%%
%2 citations counted in INSPIRE as of 22 Nov 2018





%\cite{Bergshoeff:2019pij}
\bibitem{Bergshoeff:2019pij}
E.~A.~Bergshoeff, J.~Gomis, J.~Rosseel, C.~Şimşek and Z.~Yan,
\emph{``String Theory and String Newton-Cartan Geometry,''}
J.\ Phys.\ A {\bf 53} (2020) no.1,  014001
doi:10.1088/1751-8121/ab56e9
[arXiv:1907.10668 [hep-th]].
%%CITATION = doi:10.1088/1751-8121/ab56e9;%%
%15 citations counted in INSPIRE as of 05 Mar 2020


%\cite{Kluson:2019uza}
\bibitem{Kluson:2019uza}
J.~Klusoň and P.~Novosad,
\emph{``Non-Relativistic M2-Brane,''}
JHEP {\bf 1906} (2019) 072
doi:10.1007/JHEP06(2019)072
[arXiv:1903.12450 [hep-th]].
%%CITATION = doi:10.1007/JHEP06(2019)072;%%
%3 citations counted in INSPIRE as of 09 Mar 2020

%\cite{Gomis:2019zyu}
\bibitem{Gomis:2019zyu}
J.~Gomis, J.~Oh and Z.~Yan,
\emph{``Nonrelativistic String Theory in Background Fields,''}
JHEP {\bf 1910} (2019) 101
doi:10.1007/JHEP10(2019)101
[arXiv:1905.07315 [hep-th]].
%%CITATION = doi:10.1007/JHEP10(2019)101;%%
%20 citations counted in INSPIRE as of 22 Mar 2020


%\cite{Kluson:2018vfd}
\bibitem{Kluson:2018vfd}
J.~Klusoň,
\emph{``Note About T-duality of Non-Relativistic String,''}
JHEP {\bf 1908} (2019) 074
doi:10.1007/JHEP08(2019)074
[arXiv:1811.12658 [hep-th]].
%%CITATION = doi:10.1007/JHEP08(2019)074;%%
%8 citations counted in INSPIRE as of 22 Mar 2020

%\cite{Roychowdhury:2019qmp}
\bibitem{Roychowdhury:2019qmp}
D.~Roychowdhury,
\emph{``Probing tachyon kinks in Newton-Cartan background,''}
Phys.\ Lett.\ B {\bf 795} (2019) 225
doi:10.1016/j.physletb.2019.06.031
[arXiv:1903.05890 [hep-th]].
%%CITATION = doi:10.1016/j.physletb.2019.06.031;%%
%9 citations counted in INSPIRE as of 22 Mar 2020

%\cite{Kluson:2019ifd}
\bibitem{Kluson:2019ifd}
J.~Klusoň,
\emph{``$(m,n)$-String and D1-Brane in Stringy Newton-Cartan Background,''}
JHEP {\bf 1904} (2019) 163
doi:10.1007/JHEP04(2019)163
[arXiv:1901.11292 [hep-th]].
%%CITATION = doi:10.1007/JHEP04(2019)163;%%
%13 citations counted in INSPIRE as of 22 Mar 2020



%\cite{Simon:2011rw}
\bibitem{Simon:2011rw}
J.~Simon,
\emph{``Brane Effective Actions, Kappa-Symmetry and Applications,''}
Living Rev.\ Rel.\  {\bf 15} (2012) 3
doi:10.12942/lrr-2012-3
[arXiv:1110.2422 [hep-th]].
%%CITATION = doi:10.12942/lrr-2012-3;%%
%31 citations counted in INSPIRE as of 09 Mar 2020




%\cite{Sen:2004nf}
\bibitem{Sen:2004nf}
A.~Sen,
\emph{``Tachyon dynamics in open string theory,''}
Int.\ J.\ Mod.\ Phys.\ A {\bf 20} (2005) 5513
doi:10.1142/S0217751X0502519X
[hep-th/0410103].
%%CITATION = doi:10.1142/S0217751X0502519X;%%
%453 citations counted in INSPIRE as of 09 Mar 2020

%\cite{Kluson:2018grx}
\bibitem{Kluson:2018grx}
J.~Klusoň,
\emph{``Nonrelativistic String Theory Sigma Model and Its Canonical Formulation,''}
Eur.\ Phys.\ J.\ C {\bf 79} (2019) no.2,  108
doi:10.1140/epjc/s10052-019-6623-9
[arXiv:1809.10411 [hep-th]].
%%CITATION = doi:10.1140/epjc/s10052-019-6623-9;%%
%15 citations counted in INSPIRE as of 22 Mar 2020



%\cite{Kluson:2017abm}
\bibitem{Kluson:2017abm}
J.~Kluson,
\emph{``Note about Hamiltonian formalism for Newton–Cartan string and p-brane,''}
Eur.\ Phys.\ J.\ C {\bf 78} (2018) no.6,  511
doi:10.1140/epjc/s10052-018-5993-8
[arXiv:1712.07430 [hep-th]].
%%CITATION = doi:10.1140/epjc/s10052-018-5993-8;%%
%8 citations counted in INSPIRE as of 23 Mar 2020






%\cite{Sen:1999md}
\bibitem{Sen:1999md}
A.~Sen,
\emph{``Supersymmetric world volume action for nonBPS D-branes,''}
JHEP {\bf 9910} (1999) 008
doi:10.1088/1126-6708/1999/10/008
[hep-th/9909062].
%%CITATION = doi:10.1088/1126-6708/1999/10/008;%%
%403 citations counted in INSPIRE as of 09 Mar 2020

%\cite{Bergshoeff:2000dq}
\bibitem{Bergshoeff:2000dq}
E.~A.~Bergshoeff, M.~de Roo, T.~C.~de Wit, E.~Eyras and S.~Panda,
\emph{``T duality and actions for nonBPS D-branes,''}
JHEP {\bf 0005} (2000) 009
doi:10.1088/1126-6708/2000/05/009
[hep-th/0003221].
%%CITATION = doi:10.1088/1126-6708/2000/05/009;%%
%352 citations counted in INSPIRE as of 09 Mar 2020

%\cite{Kluson:2000iy}
\bibitem{Kluson:2000iy}
J.~Kluson,
\emph{``Proposal for nonBPS D-brane action,''}
Phys.\ Rev.\ D {\bf 62} (2000) 126003
doi:10.1103/PhysRevD.62.126003
[hep-th/0004106].
%%CITATION = doi:10.1103/PhysRevD.62.126003;%%
%256 citations counted in INSPIRE as of 09 Mar 2020

%\cite{Roychowdhury:2019qmp}
\bibitem{Roychowdhury:2019qmp}
D.~Roychowdhury,
\emph{``Probing tachyon kinks in Newton-Cartan background,''}
Phys.\ Lett.\ B {\bf 795} (2019) 225
doi:10.1016/j.physletb.2019.06.031
[arXiv:1903.05890 [hep-th]].
%%CITATION = doi:10.1016/j.physletb.2019.06.031;%%
%9 citations counted in INSPIRE as of 12 Mar 2020

%\cite{Erkal:2009xq}
\bibitem{Erkal:2009xq}
D.~Erkal, D.~Kutasov and O.~Lunin,
\emph{``Brane-Antibrane Dynamics From the Tachyon DBI Action,''}
arXiv:0901.4368 [hep-th].
%%CITATION = ARXIV:0901.4368;%%
%7 citations counted in INSPIRE as of 13 Mar 2020

%\cite{Sen:2003tm}
\bibitem{Sen:2003tm}
A.~Sen,
\emph{``Dirac-Born-Infeld action on the tachyon kink and vortex,''}
Phys.\ Rev.\ D {\bf 68} (2003) 066008
doi:10.1103/PhysRevD.68.066008
[hep-th/0303057].
%%CITATION = doi:10.1103/PhysRevD.68.066008;%%
%196 citations counted in INSPIRE as of 16 Mar 2020

%\cite{Bergshoeff:2015uaa}
\bibitem{Bergshoeff:2015uaa}
E.~Bergshoeff, J.~Rosseel and T.~Zojer,
\emph{``Newton–Cartan (super)gravity as a non-relativistic limit,''}
Class.\ Quant.\ Grav.\  {\bf 32} (2015) no.20,  205003
doi:10.1088/0264-9381/32/20/205003
[arXiv:1505.02095 [hep-th]].
%%CITATION = doi:10.1088/0264-9381/32/20/205003;%%
%63 citations counted in INSPIRE as of 25 Mar 2020

%\cite{Kluson:2020aoq}
\bibitem{Kluson:2020aoq}
J.~Klusoň,
\emph{``Unstable D-brane in Torsional Newton-Cartan Background,''}
arXiv:2001.11543 [hep-th].
%%CITATION = ARXIV:2001.11543;%%
%1 citations counted in INSPIRE as of 27 Mar 2020


%%\cite{Hartnoll:2016apf}
%\bibitem{Hartnoll:2016apf}
%S.~A.~Hartnoll, A.~Lucas and S.~Sachdev,
%\emph{``Holographic quantum matter,''}
%arXiv:1612.07324 [hep-th].
%%%CITATION = ARXIV:1612.07324;%%
%%59 citations counted in INSPIRE as of 02 Nov 2017	






%%\cite{Son:2008ye}
%\bibitem{Son:2008ye}
%D.~T.~Son,
%\emph{``Toward an AdS/cold atoms correspondence: A Geometric realization of the Schrodinger symmetry,''}
%Phys.\ Rev.\ D {\bf 78} (2008) 046003
%doi:10.1103/PhysRevD.78.046003
%[arXiv:0804.3972 [hep-th]].
%%%CITATION = doi:10.1103/PhysRevD.78.046003;%%
%%623 citations counted in INSPIRE as of 02 Nov 2017

%%\cite{Balasubramanian:2008dm}
%\bibitem{Balasubramanian:2008dm}
%K.~Balasubramanian and J.~McGreevy,
%\emph{``Gravity duals for non-relativistic CFTs,''}
%Phys.\ Rev.\ Lett.\  {\bf 101} (2008) 061601
%doi:10.1103/PhysRevLett.101.061601
%[arXiv:0804.4053 [hep-th]].
%%%CITATION = doi:10.1103/PhysRevLett.101.061601;%%
%%583 citations counted in INSPIRE as of 02 Nov 2017

%\cite{Herzog:2008wg}
%\bibitem{Herzog:2008wg}
%C.~P.~Herzog, M.~Rangamani and S.~F.~Ross,
%\emph{``Heating up Galilean holography,''}
%JHEP {\bf 0811} (2008) 080
%doi:10.1088/1126-6708/2008/11/080
%[arXiv:0807.1099 [hep-th]].
%%%CITATION = doi:10.1088/1126-6708/2008/11/080;%%
%%277 citations counted in INSPIRE as of 02 Nov 2017



%%\cite{Son:2013rqa}
%\bibitem{Son:2013rqa}
%D.~T.~Son,
%\emph{``Newton-Cartan
%	Geometry and the Quantum Hall Effect,''}
%arXiv:1306.0638 [cond-mat.mes-hall].
%%%CITATION = ARXIV:1306.0638;%%
%%126 citations counted in INSPIRE as of 02 Nov 2017



%\cite{Sen:2000kd,Sen:2003bc}
\bibitem{Sen:2000kd}
A.~Sen,
\emph{``Fundamental strings in open string theory at the tachyonic vacuum,''}
J.\ Math.\ Phys.\  {\bf 42} (2001) 2844
doi:10.1063/1.1377037
[hep-th/0010240].
%%CITATION = doi:10.1063/1.1377037;%%
%111 citations counted in INSPIRE as of 24 Mar 2020

%\cite{Sen:2003bc}
\bibitem{Sen:2003bc}
A.~Sen,
\emph{``Open and closed strings from unstable D-branes,''}
Phys.\ Rev.\ D {\bf 68} (2003) 106003
doi:10.1103/PhysRevD.68.106003
[hep-th/0305011].
%%CITATION = doi:10.1103/PhysRevD.68.106003;%%
%100 citations counted in INSPIRE as of 24 Mar 2020














\end{thebibliography}
\end{document}